# Implementing Quantum Search Algorithm with Metamaterials

Weixuan Zhang,[1*] Kaiyang Cheng,[2*] Chao Wu,[2] Yi Wang,[2] Hongqiang Li,[2$] and Xiangdong Zhang[1+]

[1]Beijing Key Laboratory of Nanophotonics & Ultrafine Optoelectronic Systems, School of Physics, Beijing Institute of Technology, 100081, Beijing, China

[2]The Institute of Dongguan-Tongji University, Dongguan, Guangdong 523808; School of Physics Science and Engineering, Tongji University, Shanghai 200092, China

*These authors contributed equally to this work. [+$]To whom correspondence should be addressed. E-mail: zhangxd@bit.edu.cn; hqlee@tongji.edu.cn

**Metamaterials, artificially structured electromagnetic (EM) materials, have enabled the realization of many unconventional electromagnetic properties not found in nature, such as negative refractive index, magnetic response, invisibility cloaking and so on. Based on these man-made materials with novel EM properties, various devices have been designed and realized. However, quantum analog devices based on metamaterials have not been achieved so far. Here we designed and printed metamaterials to perform quantum search algorithm. The structures, comprising of an array of two-dimensional (2D) sub-wavelength air holes with different radii perforated on the dielectric layer, have been fabricated by using 3D printing technique. When an incident wave enters in the designed metamaterials, the profile of beam wavefront is processed iteratively as it propagates through the metamaterial periodically. After $\sim \sqrt{N}$ roundtrips, precisely the same as the efficiency of quantum search algorithm, searched**

**items will be found with the incident wave all focusing on the marked positions. Such a metamaterial-based quantum searching simulator may lead to remarkable achievements in wave-based signal processors.**

In recent years, many efforts have been devoted to the design and realization of metamaterials since the experimental demonstration of negative refraction in 2001.[1] After nearly twenty years of research, different types of metamaterials from microwave to visible frequencies have been fabricated, such as negative index metamaterials,[1–6] artificial magnetism metamaterials,[7,8] transformation optics metamaterials,[9–11] hyperbolic metamaterials,[12,13] active metamaterials,[14–16] nonlinear metamaterials,[17] quantum metamaterials[18–20] and so on.[21–27] Recently, the concept of "computational metamaterials" has been proposed.[28–33] Mathematical operations, such as spatial differentiation, integration, and convolution, can be performed by using designed metamaterial blocks. The dimension at wavelength-scale of such kind of computing machines offers the possibility of miniaturized full-wave computing systems that are several orders of magnitude thinner than conventional bulky lens-based optical processors.

On the other hand, quantum computation has been the focus of numerous studies and is expected to play an important role in future information processing, since it outperforms classical computation at many tasks.[34] The excitement over quantum computation is based on just a few algorithms, the best-known being Shor's factorization algorithm[35] and Grover's search algorithm.[36] The latter is extremely important, both from a fundamental standpoint, as it is usually more efficient than the best classical algorithm, and from a practical standpoint, because fast searching is central to solving difficult problems. Up to now, Grover's search algorithm has been implemented under many standard circuit models, such as optical experiments,[37,38] NMR systems,[39,40] trapped ion[41] and so on. The problem is whether Grover's search algorithm can be performed by using designed metamaterials? In this work, we numerically and experimentally demonstrate the implementation of Grover's search algorithm through dielectric metamaterials at microwave frequencies.

The general protocol is graphically shown in **Figure 1**a with different colors being used to distinguish the structural regions that perform different functions. In this metamaterial-based searching protocol, the electric field amplitude of the incident microwave "$E(y)$" is recognized as the probability amplitude of the equivalent quantum state, spatial positions "$y$" are used to label the items in the database, and the maximum number of the database is fixed by the full width at half maximum (FWHM) "$D$" of the incident intensity profile with near-Gaussian distribution. The designed metamaterial is comprised of four cascaded subblocks: an oracle subblock (red area, $U_m$), two Fourier transform subblocks (green areas, $F$) and a phase plate subblock (blue area, $U_0$). The role of the oracle subblock is to mark the targeted item (red arrow) |$y=m$> by imprinting a spatially-dependent phase profile on the incident beam, $E(y) \sim E(y)\exp(i2\pi\sqrt{\varepsilon_m(y)}d_m/\lambda_0)$, where $\lambda_0$ is the wavelength of the incident beam in vacuum, $d_m$ and $\varepsilon_m(y)$ are the length and effective permittivity of the subblock. Here, $0.5\pi$ phase transition is added in the narrow area around the tagged items and $0\pi$ phase transition elsewhere. A combination of two Fourier transform subblocks and the phase plate subblock can convert the phase difference marked by the oracle subblock into amplitude information by the sequences ~$FU_0F$, which is similar to the "inversion-about-average" operation (*IAA*) of the original quantum search algorithm.[36] Here, the Walsh-Hadamard transform is replaced by the Fourier transform.[37] Two

Fourier transform subblocks, which perform the Fourier transform on the optical wavefront, can be implemented by the grade-index (*GRIN*) metamaterials.[28] The corresponding permittivities of the GRIN subblocks can be expressed as: $\varepsilon_g(y) = \varepsilon_c[1-(\pi/2l)^2 y^2], |y| \leq W$, where *W* and *l* are the width and length of the GRIN block, and $\varepsilon_c$ is the permittivity at the central plane ($\varepsilon_g(y=0)$). The phase plate subblock, which imprints a phase profile in the frequency plane, ~ $\exp(i2\pi\sqrt{\varepsilon_0(y)}d_0/\lambda_0)$, can be performed by the properly designed meta-screen like the oracle subblock, except for the items with $0.5\pi$ phase transition being moved to the positions around *y*=0 (blue arrow). Similarly, $d_0$ and $\varepsilon_0(y)$ are the length and effective permittivity of the phase plate subblock, respectively. When two reflectors ($R_1$ and $R_2$) are fixed at both ends of the structure, the incident pulse will travel back and forth within the cavity, each roundtrip representing one iteration of the search algorithm. After a few iterations, the optical fields are completely concentrated on the searched positions and the marked items are proven to be found.

The above designed cascaded subblocks can be realized by drilling 2D sub-wavelength holes array with different radii on a single piece of dielectric layer based on the effective medium theory (see Supplementary Information). In this condition, the specific distribution of air-holes on each subblock can be engineered to implement its particular function. It is worth noting that our design principle is consistent with Ref. [37], where the Grover's search algorithm has been realized in lenses-based optical systems. The important difference is that all optical elements used in our design are integrated on a single metamaterials, which provide the possibility of highly compact, potentially integrable architectures within much smaller volume.

Our metamaterial is fabricated with a resolution at 0.016 mm, i.e., ~one-thousand of operational wavelength, by using 3D-printing technology, which has wide applications in many domains, such as biomedical applications and engineering.[42] Through the reasonable design based on the effective medium theory, the air holes can be fabricated precisely with 3D-printing technology, which plays a key role in implementing the searching process with quantum efficiency. The modeling material considered here is veroclear810, which is a kind of transparent material with the relative permittivity being 3.0 at the operating wavelength $\lambda_0$ =1.55 cm. The basic idea disclosed in the present paper can also be applied, in principle, to any wavelength of interest, depending on the material parameters and the dimensions. Hence, in optical and near-infrared frequencies, metamaterials with chip-scale are possible to be fabricated to complete Grover's search algorithm.

The main purpose of the paper is to perform the most basic verification of implementing Grover's search algorithm with metamaterials. Consequently, the practical metamaterial is designed to perform the simplest quantum searching algorithm with only 1.5-iterations. The more complicated metamaterial, which can perform the searching process of a larger databases, can also be designed using the same method (see numerical results in Supporting Information). While, the more sophisticated machining technology, higher detection precision and high frequency ultrashort pulse are needed experimentally. Figure 1b presents the photograph of the printed metamaterial (3-cm-thick) with the corresponding parameters being: $\varepsilon_c = 3.0$, $d_m$=$d_0$=3$a$, $W$=51$a$ and $l$=57$a$, respectively. Here, $a$ (~0.4 cm) is the size of the unit cell that divides our metamaterial system into a series of small square grids. The designed GRIN block consists of 51 by 57 holes, and each hole occupies a 0.4 by 0.4 cm² square cell. The largest radius at the position of $y = \pm 10 \text{ cm}$ is 0.1904 cm, and the smallest radius at $y$=0.4 cm is 0.0076 cm (0 cm at

$y=0$). In the case of the $U_{0(m)}$ block, it is composed of 51 by 3 holes, and each hole occupies the same square cell as the GRIN block. The radius at the position with $0.5\pi$ phase transition (from -5a to -3a, blue arrow) is 0.1 cm, and the radius at the position with $0\pi$ phase transition is 0.1841 cm. In order to reduce the influence of the optical diffraction, the gradually varied radius (0.118 cm, 0.134 cm, 0.148 cm, 0.161 cm, 0.173 cm) is used between the positions of the targeted items and surrounding areas. Moreover, the designed metamaterial can also perform multi-items search, when the oracle subblock possesses more than one marked positions. In this regard, we designed the second metamaterial with the tagged items lying at two different locations (blue arrows), shown in Figure 1c. When the searching process is completed, the incident microwaves will focus on these two-tagged positions.

In order to test the efficiency of the searching scheme, we perform full-wave numerical simulations using finite element method (COMSOL Multiphysics). **Figure 2**a shows the snapshot of the electric field intensity throughout the first metamaterial with the input function being $E(y)=\exp(-y^2/20)$. In this case, the *FWHM* value of the incident intensity equals to $D$=5.26 cm. In Figure 2b, we also plot the distribution of the wave intensity $|E(y)|^2$ at the output plane with 0.5-iteration. It is found that the incident wave is mainly focused on the position around the marked holes. However, there are still a few of the electric fields not concentrated on the searched positions. With the increasing of the iterations, the concentration of electric fields will become more significant. This is equivalent to the condition that the probability of the searched state increases continuously (approaches to unity) with the increment of the searching time serially. Figure 2c presents the distribution of the output intensity with the beam propagating 1.5-roundtrips within the metamaterial. It is clearly shown that the output microwaves are nearly all concentrated on the marked positions with the corresponding *FWHM* being *d*=1.4 cm. We need to emphasize that the *FWHM* value of the output intensity is not equal to the real size of the marked space, due to the inhomogeneous distribution of the incident beam. Several little side peaks are result of the diffraction effect in the process of iterations. The corresponding snapshot of the field intensity is plotted in Figure 2d. The ratio *D/d* can be interpreted as the size of the database (*N*) for a single item search. A straightforward calculation demonstrates that the number of iterations performed on the metamaterial is precisely consistent with the efficiency of the quantum search algorithm $\pi\sqrt{D/d}/4 \approx 1.5$,[36] when the marked items are located. It indicates that the designed metamaterial is capable of implementing the quantum search algorithm. The phenomenon originates from wave interference properties, entanglement is not necessary for the efficiency of quantum searching arithmetic, as noted in Refs. [37, 38].

The snapshot of the electric field intensity throughout the second metamaterial (with two marked items) is shown in Figure 2e. The output profile is presented in Figure 2f. We can see that only after 0.5-iteration, the incident electric fields are nearly all concentrated on the two marked spaces. In fact, multi-iterations searching processes (such as 2.0-iterations) can also be realized in such a case, which will exhibit $\sim \sqrt{N}$ scaling behaviors as expected from Grover's algorithm. The detailed results are provided in Supplementary Information.

To further validate our design, experimental measurements are carried out. The conceptual drawing of the experimental setup is illustrated in **Figure 3**a. In experiments, it is possible to reduce the microwave scattering to two dimensions by confining electromagnetic waves between two parallel conducting planes. The microwave cavity is surrounded by the absorbing materials to reduce the experimental errors induced by the optical scatterings. A beam with a near-Gaussian distribution in wavefront can be generated by a standard horn antenna at Ku band. The pulse signals, whose carrier frequency and period are 19.3 GHz and 100 ns, are generated by the arbitrary waveform generator (Keysight M9502A AxleChassis) with the pulse width, 1 ns, being short compared to the roundtrip time (about 2.87 ns), and so that the pulse signals belonging to different iterative times can be resolved by the oscilloscope (Keysight DS0-Z 254A) with the sampling frequency being 80GHz. And a microwave amplifier (Keysight 83017A) is also used to ensure a big enough signal input for detection. In addition, in order to perform the searching process iteratively, we install two ceramic ($Al_2O_3$) sheets on both ends of the metamaterial to make the incident pulse travel back and forth within the cavity. Figure 3b presents the photograph of the ceramic sheet. The reflectivity of the ceramic sheet is about 55% with the corresponding thickness being 1 mm (see Supplementary Information). In Figures 3c and d, we plot two typical time traces with the probing pin (monopole antenna fixed on the 3D scanner) locating at $y$=-1.4 cm and $y$=1.4 cm, respectively. Three discrete pulses in time domain, each with a widths about 1 ns, are clearly presented, which correspond to the searching signals for 0.5-, 1.5- and 2.5-iterations, respectively. The interval of the adjacent signals are both nearly 2.87 ns equaling to the time of the pulse travelling back and forth between the cavity mirrors. We can see that the intensities at $y$=-1.4 cm (marked position) are always larger than that at $y$=1.4 cm. This is consistent with the fact that the incident microwaves are focused at the searched position and accumulated in time. We measured the entire beam profile by recording traces, just as in Figure 3c and 3d, at different positions. It is noted that the bandwidth of the 1ns pulse signal is about 1GHz. Consequently, the Fourier spectral analysis (FSA) is needed to get the intensity of a specific frequency. Combining intensities on different positions, the transverse beam profile of the output signal can be drawn. The evolution of such profiles of different roundtrips shows how the algorithm proceeds.

As shown in Figure 3e and 3f, the experimental measured output profiles with 0.5- and 1.5-iterations are nearly consistent with the corresponding numerical simulations. The wider peak compared to the numerical solution is the result of the imperfect sample fabrication and minor variation for the incident wave profile compared to the ideal condition. For example, it is very difficult to print perfect air holes with small radii on the veroclear810 layer. On the other hand, the multi-items searching process with the other sample is also measured. As shown in Figure 3g, the incident microwaves are almost concentrated on the two marked positions with 0.5-iteration. The results of multiple-tests are consistent. The experimental demonstration indicates that the quantum search algorithm can be simulated successfully using the designed metamaterials, both for cases with single and two marked items.

In summary, we have proposed and designed metamaterials to perform wave-based quantum search algorithm. To the best of our knowledge, no studies have dedicated on this topic before. Our numerical simulations and experiments clearly confirm that the searching efficiency, on such metamaterials fabricated by 3D printing technology, is the same as quantum computing. Moreover, the general design principle in our device can be applied, in principle, to any wavelength. Apart from implementing Grover's search algorithm, other quantum algorithms can also been simulated based on the similar methods, such as Deutsch-Jozsa algorithm. Our metamaterial approach for an all-optical quantum searching simulator provides a new way to shed light on quantum analog behaviors, which may lead to remarkable achievements in wave-based signal processors.

**Experimental Section**

3D printing is utilized for the rapid prototyping of 3D models originally generated by a computer aided design program. The 3D printer (Objet 1000 plus) is controlled with the computer program (Objet studio) in this work. The printing temperature and humidity are 20° C and 54%, respectively. The temperature of the printing head is 67° C. The support materials 249g sup705 and modeling materials 3578g veroclear810 are used. The printing time is 13.27 hours. After the completion of the printing process, the support materials are needed to be clear away (water cannons flush tiny support materials) and the modeling materials are needed to be baked. The baking temperature and baking time are 50° C and 10 hours, respectively.

**Supporting Information**
Supporting Information is available from the Wiley Online Library or from the authors.

**Acknowledgements**
This work was supported by the National Natural Science Foundation of China (Grant No. 11574031 and 61421001). W. Zhang and K. Cheng contributed equally to this work.

Received: ((will be filled in by the editorial staff))
Revised: ((will be filled in by the editorial staff))
Published online: ((will be filled in by the editorial staff))

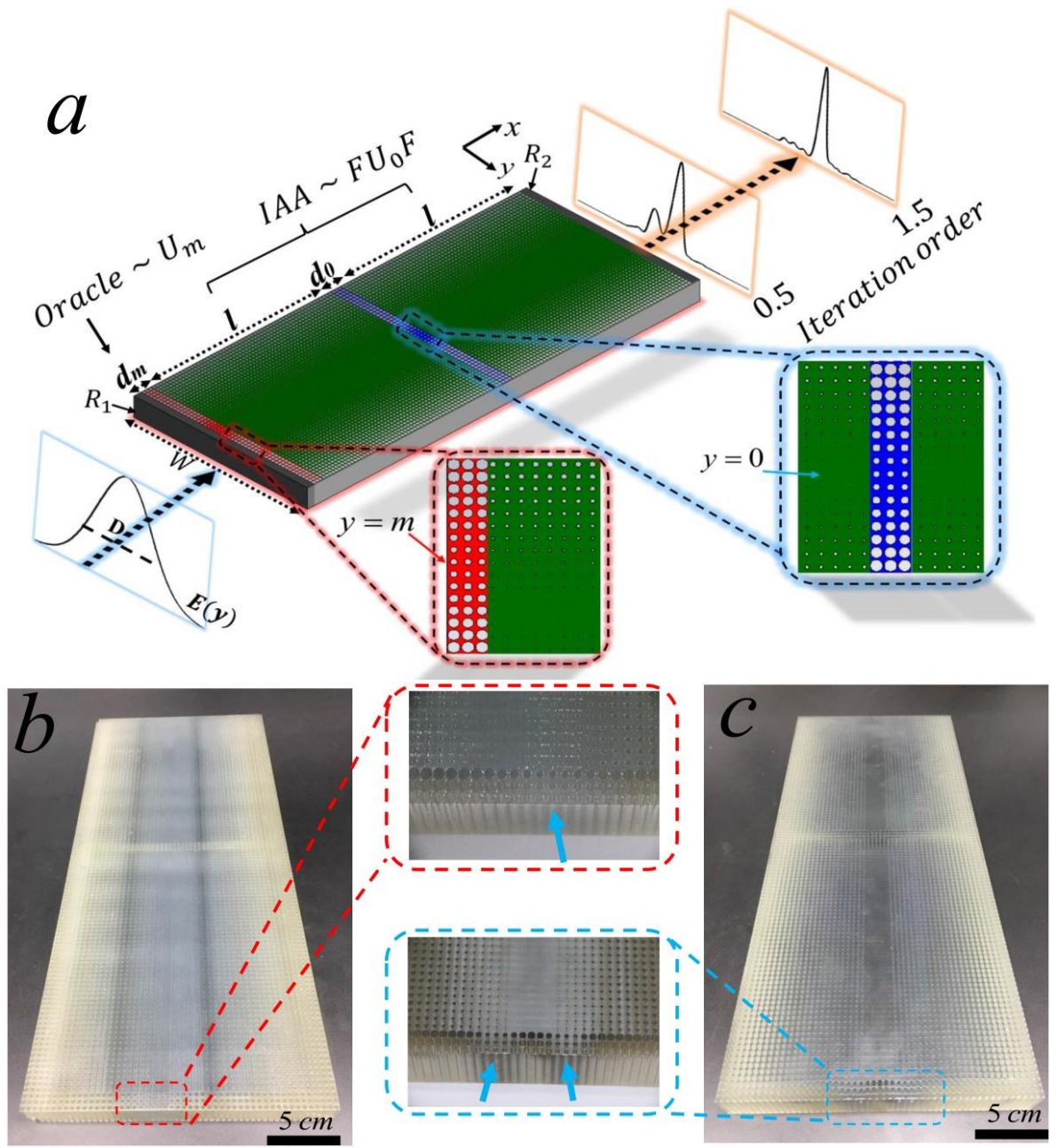

**Figure 1.** Schematic of the metamaterials implementing quantum search algorithm. a) The general protocol of simulating quantum searching arithmetic with metamaterial. b,c) The photographs of the first and second metamaterial samples fabricated by using 3D printing technology. The two amplifying images plot the marked positions of oracle blocks. In order to reduce the influence of the optical diffraction, a gradually varied phase profile is used between the positions of the targeted items (0.5$\pi$ phase) and surrounding areas (0$\pi$ phase).

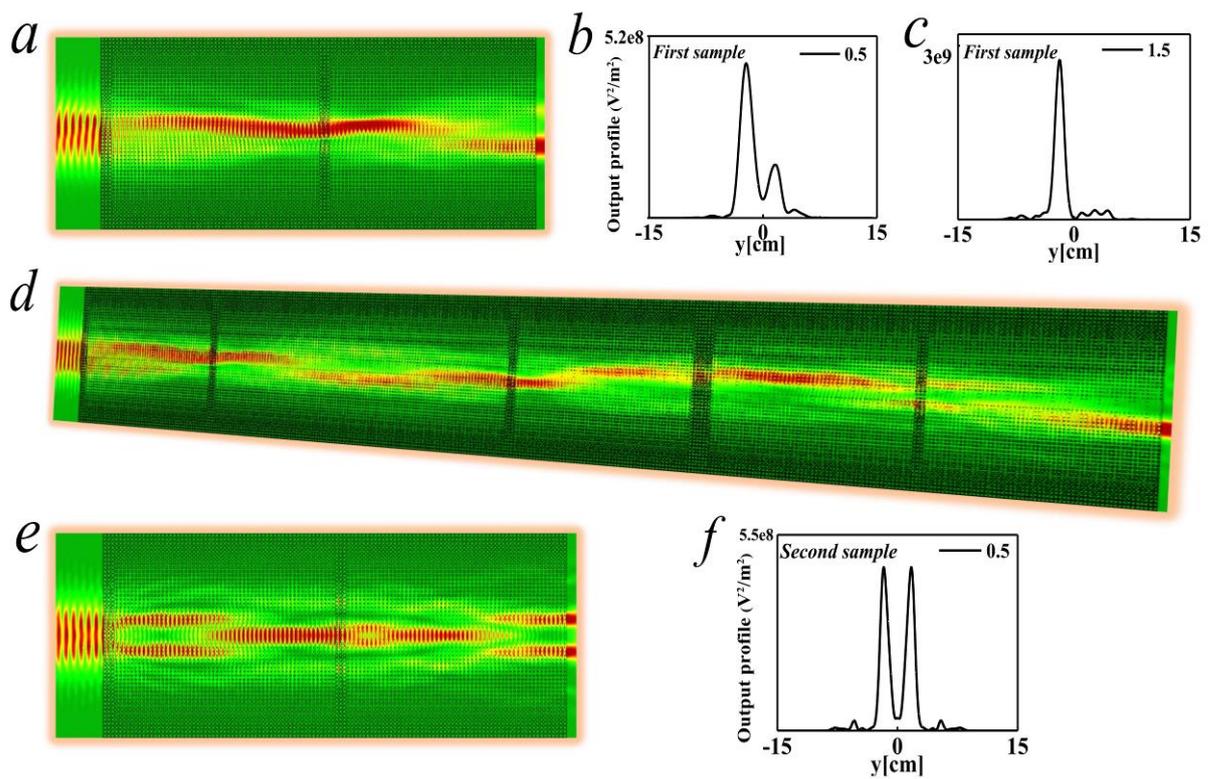

**Figure 2. Theoretical simulation of the quantum search algorithm.** a,d) The snapshots for the intensity of the incident microwave throughout the first metamaterial functioning as quantum searching simulator with 0.5- and 1.5-iterations, respectively. b,c) Simulation results for the output intensity of the first metamaterial with the incident wave propagating 0.5- and 1.5-roundtrips within the metamaterial, respectively. e) The snapshot for the intensity of the incident wave throughout the second metamaterial with 0.5-iteration. f) Simulation results for the output intensity of the second metamaterial with the incident wave propagating 0.5-roundtrip within the metamaterial.

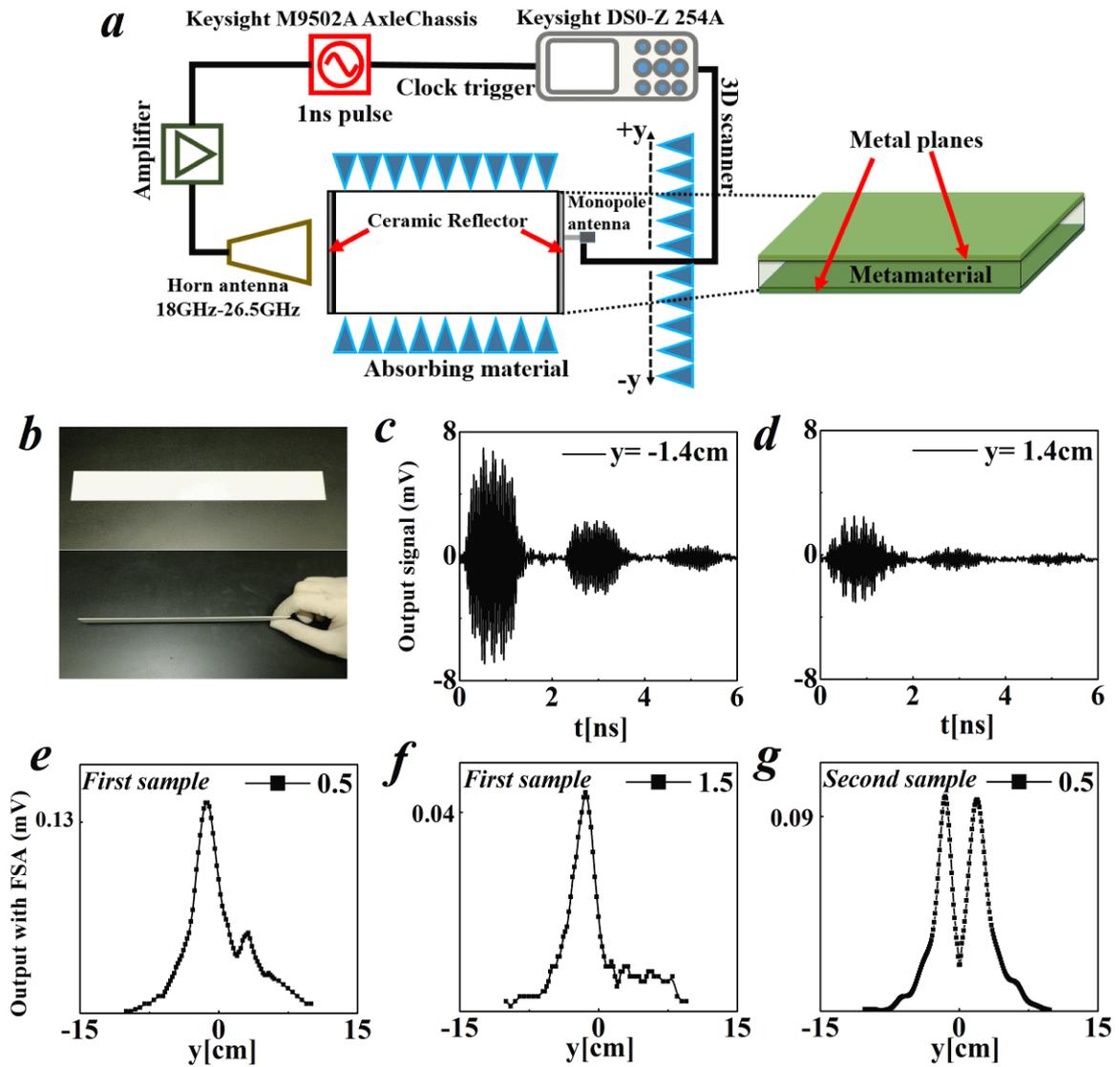

**Figure 3. Experimental measurements of the quantum search algorithm.** a) The conceptual drawing of the experimental setup. b) Photograph of the ceramic reflector. c) and d) Time traces of the output intensities for the first metamaterial at positions of *y*=-1.4 cm and *y*=1.4 cm, respectively. e,f) Experimental results for the profiles of the output intensity of the first metamaterial with 0.5- and 1.5-iterations, respectively. g) Experimental result for the profile of the output intensity of the second metamaterial with 0.5-iteration.

# Supporting Information

**Implementing quantum search algorithm with metamaterials**

*Weixuan Zhang, Kaiyang Cheng, Chao Wu, Yi Wang, Hongqiang Li\* and Xiangdong Zhang\**

## *1. The protocol of implementation of Grover's search algorithm with metamaterials.*

A quantum state with uniform superposition is prepared for Grover's search algorithm. A quantum operation called Grover's iterator is then applied repeatedly to it. This operation does the following: first the amplitude of a single tagged state is inverted by using the oracle operator. The second step is an inversion of all amplitudes about their average value, which is realized by "iversion-about-average" (IAA) operation. By repeating the Grover iterator $\sim \sqrt{N}$ times, we gradually transform the state from the uniform superposition into the pure tagged state and we thus find our solution.

The Grover's search algorithm maps onto our metamaterial-based protocol as follows. The electric field amplitude of the incident microwave is recognized as the probability amplitude of the equivalent quantum state, spatial positions are used to label the items in the database, and the maximum number of the database is fixed by the full width at half maximum (FWHM) of the incident intensity profile with near-Gaussian distribution. The function of oracle block, which marks the targeted item by imprinting a spatially-dependent phase profile ($0.5\pi$ in a narrow marked area and 0 $\pi$ elsewhere) on the incident beam, is equivalent to the oracle operator in Grover's search algorithm. Then, a combination of two Fourier transform subblocks and the phase plate subblock can convert the phase difference marked by the oracle subblock into amplitude information by the sequences ~$FU_0F$, which is similar to the *IAA* operation of the original quantum search algorithm. When an incident wave enters in the designed metamaterials, the profile of beam wavefront is processed iteratively as it propagates through the metamaterial periodically. After ~$\sqrt{N}$ iterations, which is the same as the efficiency of quantum search algorithm, our designed metamaterial will transform the incident waveform with near-Gaussian distribution into the bright spot focused on the marked positions.

## *2. Designing metamaterials based on the effective medium theory.*

Using the following formula, which is valid in the quasi-static regime, we can write the effective permittivity $\varepsilon_{eff}$ of the metamaterial as: [S1]

$$\varepsilon_{eff} = f_{host}\varepsilon_{host} + f_{air}\varepsilon_{air}, \tag{S1}$$

where $f_{air}$ and $f_{host}$ are the volume fractions of the air hole and dielectric layer, respectively. On the basis of our design, the volume fraction of the air hole can be expressed as:

$$f_{air}(x, y) = \pi r(x, y)^2 / a^2, \tag{S2}$$

where $r(x, y)$ is the spatially varying radius of the air holes. It is noted that the effective medium theory is valid in the long-wavelength limit, so the parameter *a* should be much less than the wavelength of the incident light in vacuum to avoid the Bragg reflection. [S2] Combing Equation S1 and S2, $r(x, y)$ can be determined using the following formula:

$$r(x, y) = a[(\varepsilon_{eff}(x, y) - \varepsilon_{host}) / (\pi(\varepsilon_{air} - \varepsilon_{host}))]^{1/2}. \tag{S3}$$

On the other hand, the effective permittivities of the $U_m$ ($U_0$) and GRIN subblocks can be

expressed as:

$$\varepsilon_{m(0)}(y) = \begin{array}{ll} (\dfrac{1.25\lambda_0}{d_{m(0)}})^2, & 0.5\pi \\ (\dfrac{\lambda_0}{d_{m(0)}})^2, & 0.0\pi. \end{array} \quad for\ U_{m(0)}$$

$$\varepsilon_g(y) = \varepsilon_c[1-(\pi/2l)^2 y^2], \quad for\ GRIN. \tag{S4}$$

Combing Equation S3 and S4, the whole metamaterial can be realized by drilling sub-wavelength holes array on a single piece of dielectric layer. The holes sizes and distributions are different on the four cascaded subblocks, which results from the different values of the effective permittivities $\varepsilon_{eff}$ at different positions.

## 3. *Simulation methods.*

COMSOL Multiphysics, version 5.2, is used for all simulations presented in the main text and Supplementary Information. The schematic model used in the simulation for the matamaterials (0.5-iteration) with the two-dimensional GRIN blocks and $U_{0(m)}$ blocks is shown in Figure S1. The whole metamaterial ($U_0$/GRIN/$U_m$/GRIN) presented here is limited by the width *W* in the transverse direction *y*, and by the total length $d_0+l+d_m+l$ in the longitudinal direction *x*. Outsides the metamaterial, the metarial is defined as air (green area), and each side air region has a width of *W/4*. The left boundary of air domain is the incident port (black dash line). Extra perfectly-matched-layer regions (blue area) are used around the whole structure with thickness $\lambda_0/4$. The external boundaries are defined as scattering boundary condition. The maximum mesh size is 0.1 cm, which is fine enough for convergence.

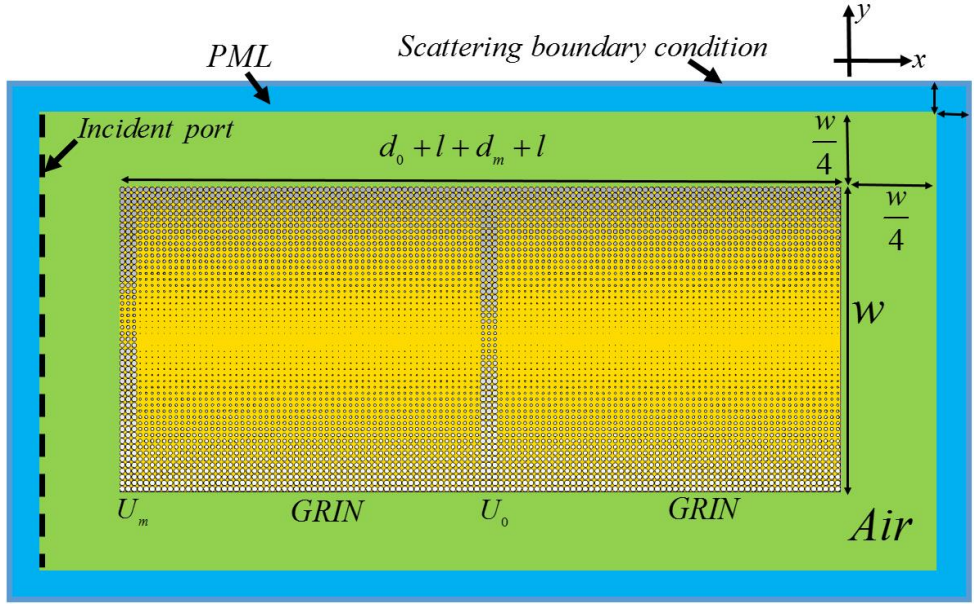

**Figure S1.** Basic model built in COMSOL.

## 4. Single-item searching with multiple iterations

The more complicated metamaterial (compared to the case mentioned in the main text), which can simulate the searching process of a larger databases, can also be designed. Here, we present the simulation results with the marked items being found after 2.5-iterations. The parameters are chosen as: $\varepsilon_c = 8.0$, $d_m=d_0=3a$, $W=100a$ and $l=120a$, respectively. The marked items are located on the positions from $y=-5a$ to $-2a$. The input function is expressed as: $E(y)=exp(-y^2/50)$ with the corresponding *FWHM* value of the incident intensity equaling to $D=10$ cm. Figure S2 shows the distributions of the light intensity $|E(y)|^2$ at the output plane with 0.5 (black line), 1.0 (red line), 1.5 (blue line), 2.0 (green line), 2.5 (pink line), 3.0 (yellow line) and 3.5 (purple line) iterations, respectively. We can see that output lights are nearly all focused on the marked positions after 2.5-iterations with the total *FWHM* being d=1.19 cm. The number of iterations performed on the metamaterial is identical to the efficiency of the quantum searching arithmetic $\pi\sqrt{D/d}/4 \approx 2..$

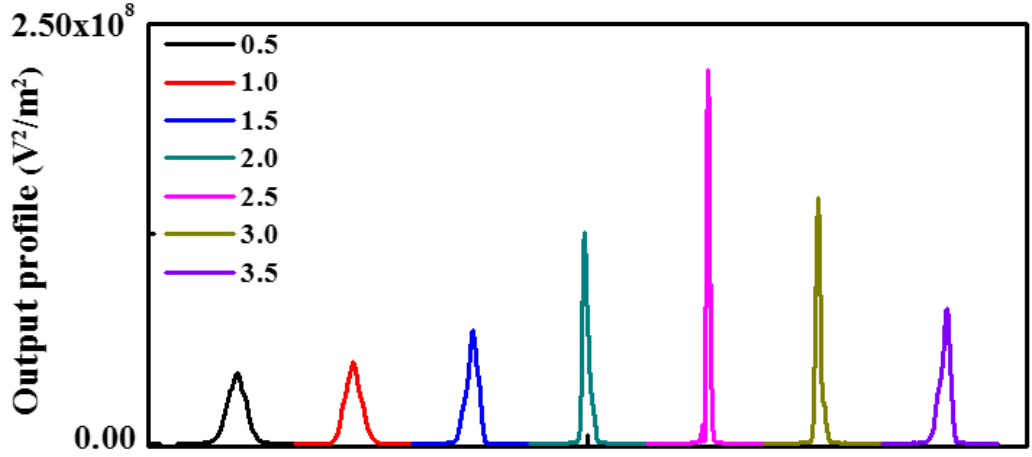

**Figure S2.** Simulation results for the output intensity with the incident light propagating 0.5 (black line), 1.0 (red line), 1.5 (blue line), 2.0 (green line), 2.5 (pink line), 3.0 (yellow line) and 3.5 (purple line) roundtrips within the designed metamaterial, respectively.

## 5. *Multi-items searching with multiple iterations*

Here, we will present the numerical results of the multi-items searching process with the incident light propagating multiple roundtrips within the designed metamaterial. The parameters are chosen as: $\varepsilon_c = 3.0$, $d_m=d_0=3a$, $W=153a$ and $l=171a$, respectively. The marked items are located on the positions from $y=-14a$ to $-8a$ and $y=8a$ to $14a$. The input function is expressed as: $E(y)=exp(-y^2/250)$. The corresponding *FWHM* value of the incident intensity equals to $D=20.5$ cm. Figure S3 plots the distributions of the light intensity $|E(y)|^2$ at the output plane with 0.5 (black line), 1.0 (red line), 1.5 (blue line), 2.0 (green line) and 2.5 (pink line) iterations, respectively. It is clearly shown that output lights are nearly all concentrated on the two marked positions after 2.0-iterations with the total *FWHM* being $d=3.43$ cm. In this case, the number of iterations performed on the metamaterial is consistent with the efficiency of the quantum searching arithmetic $\pi\sqrt{D/d}/4 \approx 2$. Consequently, we demonstrate that the designed metamaterial can also implement the multi-items searching process with quantum efficiency.

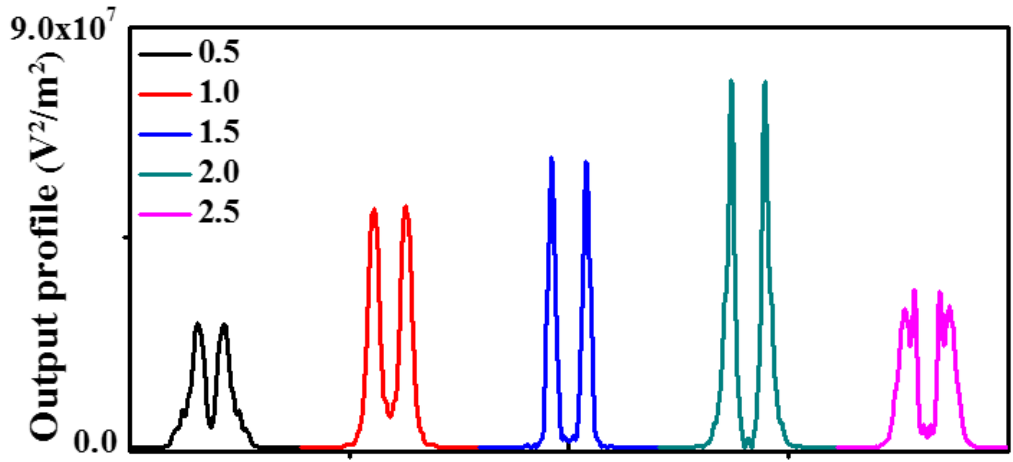

**Figure S3.** Simulation results for the output intensity with the incident light propagating 0.5 (black line), 1.0 (red line), 1.5 (blue line), 2.0 (green line) and 2.5 (pink line) roundtrips within the designed metamaterial, respectively.

## 6. Designing the ceramic ($Al_2O_3$) reflector.

The thickness of the $Al_2O_3$ sheet has a remarkable influence on the corresponding optical response. Figure S4a shows the transmittance (black line) and reflectance (red line) as functions of the thickness of $Al_2O_3$ sheet, when the frequency of the incident light is 19.3GHz. At this frequency, the relative permittivity of $Al_2O_3$ is 8.2. The peaks and valleys of the transmittance and reflectance repeat periodically with the variation of the thickness of the sheet. Within the machining precision, the 1 mm-thick $Al_2O_3$ sheets are chosen to be fabricated. In this condition, the multiple reflection within the $Al_2O_3$ sheet can be ignored. Figure S4b presents the transmittance and reflectance as functions of the incident light frequency, when the thickness of the $Al_2O_3$ sheet is 1 mm. We can see that changes of the transmittance and reflectance are negligible in the bandwidth of the incident pulses.

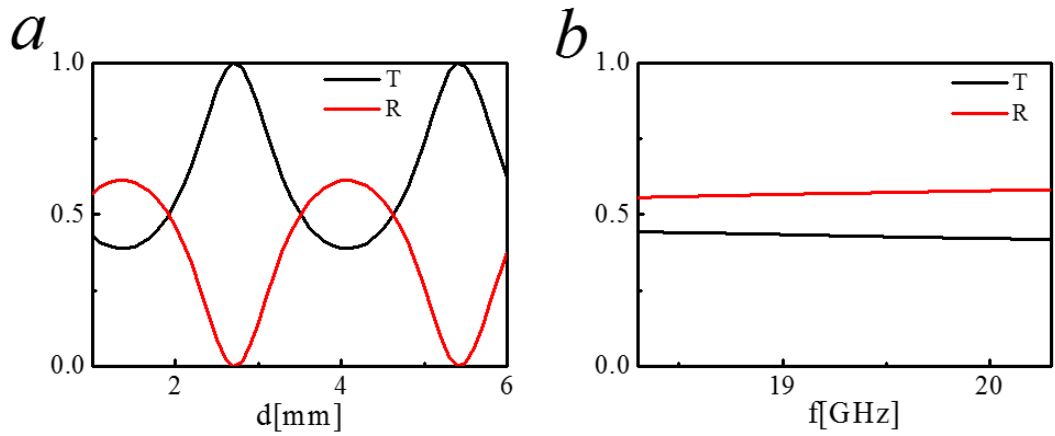

**Figure S4.** a) The relationship between the thickness and transmittance (reflectance) of the ceramic sheet with the incident frequency being 19.3GHz. b) The relationship between the incident frequency and transmittance (reflectance) with the thickness of the ceramic sheet being 1 mm.